\documentclass[11pt,twoside]{article}
\usepackage{asp2004}
\usepackage{psfig}
\usepackage{epsf}
\usepackage{graphics}
\usepackage{lscape}
\def\edcomment#1{\iffalse\marginpar{\raggedright\sl#1\/}\else\relax\fi}
\reversemarginpar

\newcommand{\ApJS}{\apjs}
\newcommand{\ApJL}{ApJL}
\newcommand{\ApJ}{\apj}
\newcommand{\PRL}{\prl}
\newcommand{\PRD}{\prd}
\newcommand{\MNRAS}{\mnras}

\newcommand{\etal}{{\it et al.}}
\newcommand{\aut}[2]{{#1, #2.,}}
\newcommand{\paut}[2]{{#1, #2.,} \&}
\newcommand{\laut}[2]{{#1, #2.,}}
\newcommand{\refs}[6]{#5, #2, #3, #4}

\newcommand{\mybib}[2]{\bibitem[\protect\astroncite{#1}{}]{#2}}
\newcommand{\submitted}{{\rm submitted}}
\newcommand{\inpress}{{\rm in press}}
\def\simlt{\la}

\begin{document}
\title{Dark Energy Probes in Light of the CMB}
\author{Wayne Hu}
\affil{Kavli Institute for Cosmological Physics, Enrico Fermi Institute and Department of Astronomy
and Astrophysics, Chicago IL 60637 USA}

\begin{abstract}
CMB observables have largely fixed the expansion history of the universe in the deceleration
regime and provided two self-calibrated absolute standards for dark energy
studies: the sound horizon at recombination as a standard ruler and the amplitude of
initial density fluctuations.  We review these inferences
and expose the testable assumptions about recombination and reionization 
that underly them.  Fixing the deceleration regime with CMB observables, deviations in the
distance and growth observables appear most strongly at $z=0$ implying that
accurate calibration of local and CMB standards may be more important than redshift 
range or depth.  The
single most important complement to the CMB for measuring the dark energy equation
of state at $z \sim 0.5$ is a determination of the Hubble constant to better than a few percent.  
Counterintuitively, with fixed fractional distance errors and relative standards such as SNe,
the Hubble constant measurement is best achieved in the high redshift deceleration regime.
Degeneracies between the evolution and {\it current} value of the equation of state or
between its value and spatial curvature can be broken if percent level measurement and
calibration of distance standards can be made at intermediate redshifts or the growth
function at {\it any} redshift in the acceleration regime.   We compare 
several dark energy probes available to a wide and deep optical survey: baryon features
in galaxy angular power spectra and the growth rate from galaxy-galaxy lensing, shear
tomography and the cluster abundance.
\end{abstract}
\thispagestyle{plain}

\section{Introduction}

Originating mainly from high redshift, CMB observables provide few direct constraints
on the dark energy.  Nonetheless, their indirect impact on other more local dark energy probes
is important to bear in mind when planning future studies.  CMB observables provide two
things for dark energy probes: internally or self-calibrated standard rulers and fluctuations
for distance and growth rate probes and an expansion history for the universe that is
fixed as a function of redshift beyond the current acceleration regime.

We begin in \S \ref{sec:parameterization} by reviewing the dark energy observables themselves. 
In \S \ref{sec:CMB}, we discuss the calibration of the CMB standards and the means
by which their self-consistency may be checked.  We illustrate these considerations with the recombination calculation.   
Although the state-of-the-art in recombination  is sufficient
 compared to current measurement errors, it will introduce substantial systematic
errors for experiments that are cosmic variance limited out to multipoles of $\ell \sim 10^3$ like
Planck.  Errors in the assumed recombination, including exotica such as a variation in 
the fine structure constant, would appear as an inconsistency in the damping tail.  
Accepting the standard thermal history the expansion history is fixed in the deceleration
regime.  In \S \ref{sec:deviations} we explore the implications for dark energy probes.
In light of the CMB, the focus returns to low redshift dark energy observables and
accurate local calibration.
In \S \ref{sec:forecasts}, we consider several optically based probes of the dark energy
in light of the CMB.

\section{Standard Parameterizations}
\label{sec:parameterization}

On scales well below current horizon or Hubble distance where the dark energy
can be taken to be smooth, all of its observable effects come through the evolution of
its average density $\rho_{\rm DE}(a)$ as a function of
 scale factor $a = (1+z)^{-1}$.  This evolution in turn is related 
 by energy conservation
\begin{equation}
\rho_{\rm DE}(a) = \Omega_{\rm DE}\rho_{\rm crit}\Big|_{a=1} e^{-3{\int (1+w) d\ln a}}\,,
\end{equation}
 to
its current energy density $\Omega_{\rm DE}$ relative to critical 
$\rho_{\rm crit}(a) = {3 H(a)^2 / 8\pi G}$
and its
equation of state 
\begin{equation}
p_{\rm DE}(a) = w(a) \rho_{\rm DE}(a)\,.
\end{equation}
Hereafter when an argument to 
$\Omega_{\rm DE}$ and $\rho_{\rm crit}$ is unspecified, $a=1$ is to be understood.  We employ
throughout units where $c=\hbar=1$.

The two basic dark-energy dependent observables are distance and growth rate.
Distance measures are based on having 
standardized candles, rulers, or object number densities
as a function of redshift; growth rate measures are based on  standard
density fluctuations in linear theory either calibrated today or
at the initial conditions and then observed at different redshifts. 
All distance measures are ultimately based on the comoving distance to redshift $z_i$
\begin{equation}
D_i \equiv \int_{a_i}^1 { d a \over a^2 H(a) } = \int_0^{z_i} { dz \over H(z)}\,.
\end{equation}
For example, the physical angular diameter distance and luminosity
distance are related to 
 the comoving angular diameter distance 
\begin{equation}
{\cal D}_i = R \sin( D_i / R)
\end{equation} 
by multiplying and dividing by $a_i$ respectively.  Here
  $R =  H_0^{-1} (\Omega_{\rm T} - 1)^{-1/2}$ is the radius of curvature in an
open or closed geometry and $\Omega_{\rm T} \approx \Omega_m + \Omega_{\rm DE}$ is the
total energy density relative to critical.  For comoving
distances much smaller than the 
curvature scale ${\cal D}_i \rightarrow D_i$.
The
comoving volume element employed in number density tests
is $d V =  {\cal D}_i^2 dD d\Omega = {\cal D}_i^2 (dz/H_i)  d\Omega$
where $d\Omega$ is the solid angle.  If the absolute brightness or physical scale of the standards
is unknown then a $z_0 \rightarrow 0$ vs $z_i$ comparison will yield
the distance ratio
\begin{equation}
{ {\cal D}_i \over {\cal D}_0 } z_0 \rightarrow H_0 {\cal D}_i \,,
\end{equation}
i.e. a distance measured in units of $h^{-1}$ Mpc.  This distinction differentiates (absolute) CMB
calibration standards from (relative) local standards.  Likewise, if the standard ruler is
employed in the radial or redshift direction, it measures $d D/dz = H^{-1}$ if absolutely calibrated
and $H_0 dD/dz = (H_0/H)$ if relatively calibrated.  Finally the ruler need not even be standard in redshift if
 its angular and radial extent are compared at the same redshift
 (\cite{AlcPac79} 1979).  In
this case the quantity measured is $H {\cal D}$.

Under the assumption that the density perturbations are dominated by fluctuations
in the non-relativistic matter $\delta_m \equiv \delta\rho_m/\rho_m$,  they
evolve
under self-gravity as 
\begin{equation}
{d^2 \delta_m  \over dt^2} + 2H(a) {d \delta_m \over dt} = 4\pi G \rho_m(a) \delta_m \,.
\label{eqn:deltadot}
\end{equation}
The growth rate is more usefully represented in terms of the scale factor and relative
to the rate during the matter dominated epoch.  Defining the growth
rate $G(a) \propto \delta_m/a$, Eqn.~(\ref{eqn:deltadot}) becomes
\begin{eqnarray}
\frac{d^2 G}{d\ln a^2} + \left( 4 + {d \ln H \over d\ln a} \right) \frac{d G}{d\ln a} 
+ \left[ 3+ {d \ln H \over d\ln a}  - {3\over 2}\Omega_m(a) \right] G = 0 \,.
\end{eqnarray}
For the growing mode of density perturbations one solves this equation with initial conditions
of $G=1$ and $d G /d\ln a=0$.
Under the assumption of a flat universe, $\Omega_m(a) +\Omega_{\rm DE}(a)=1$, $G$
is solely a function of the dark energy density
\begin{eqnarray}
\frac{d^2 G}{d \ln a^2}  + \left[ \frac{5}{2} - \frac{3}{2} w(a) \Omega_{\rm DE}(a) \right]
\frac{d G}{d \ln a}  + 
\frac{3}{2}[1-w(a)]\Omega_{\rm DE}(a) G =0\,.
\label{eqn:growth}
\end{eqnarray}
During the matter dominated epoch $\Omega_{\rm DE} \rightarrow 0$ and $G = $ const.\ solves
the equation of motion.
Eqn.~(\ref{eqn:growth}) is actually correct in general relativistic perturbation theory 
out to the sound horizon of the dark energy with the generalization
that $G$ is the decay rate of the gravitational potential $\Phi \propto G$ in Newtonian gauge 
(\cite{EisHu99} 1999).   

A completely empirical description of the dark energy would require that observations
constrain a free function $w(a)$.  
 However, the dark energy
will typically only have observable effects during the recent few e-foldings  when
it contributes substantially
to the expansion rate.  During this short period in the expansion history, the equation of
state function can be usefully approximated through a linear expansion around
a normalization epoch $a_n$ through the local slope  $w_{a}= - dw/da |_{a_n}$ as
\begin{equation}
w(a) = w_{n} + (a_{n}- a) w_a\,.
\label{eqn:wn}
\end{equation}  
As pointed out by \cite{Lin03} (2003), this
expansion is more stable to extrapolation away from $a_n$ than the analogous
linearization in redshift.    A common choice for the normalization
point in the literature is the present epoch  $a_{n}=1$.  In this case, let us define
the amplitude $w_{0} \equiv w_{n}$, the equation of state parameter today.
While we follow this 
convention here, it is well known that this choice will cause a  degeneracy
between the amplitude $w_{0}$ and evolution $w_{a}$ of $w(a)$ for
typical observables. Thus when marginalized over $w_a$, the parameter $w_0$
will have large errors.  Large errors does {\it not} mean that
that the equation of state parameter is nowhere well-determined, 
it simply means that given the possibility of evolution its {\it current}
value is not well determined.  Note that a measurement of $w \ne -1$ at 
{\it any} redshift would rule out a cosmological constant as the dark energy.

A given set of phenomena will best constrain $w$ at the redshifts relevant
to the phenomena. It would thus be better to place $a_{n}$ near the expected
epoch of dark energy domination.  
This situation is completely analogous to choosing a scale for the
normalization of the power spectrum, e.g. one would not quote a
COBE-style horizon scale normalization at $k = H_0$ for the WMAP data 
but a much smaller scale of $k \sim 0.02-0.1$ Mpc$^{-1}$
corresponding to the scale of the acoustic peak measurements
[see Eqn.~(\ref{eqn:initialpower})].

As with the power spectrum normalization, an inappropriate
choice for the dark energy normalization point can be corrected
after the fact {\it if} the covariance matrix of the parameters is also given. 
Suppose one analyzes the data with a dark energy parameterization of
$p=(w_0, w_a)$ then the covariance matrix of an alternate choice
$p'=(w_n, w_a)$ is given by the Jacobian transformation  
\begin{equation}
C_{\mu\nu}' = \sum_{\alpha\beta} 
{\partial p_\alpha' \over \partial p_\mu} 
C_{\alpha\beta}
{\partial p_\beta' \over \partial p_\nu} \,.
\label{eqn:rotation}
\end{equation}
In particular, there is a ``best" choice for a given observation
(\cite{EisHuTeg99b} 1999b; \cite{HuJai03} 2003)
\begin{equation}
{a_n}={a_{\rm pivot}} \equiv {C_{w_{0}w_a}\over 
C_{w_a w_a}}+1\,,
\label{eqn:pivot}
\end{equation}
such that the errors on $w_{\rm pivot}$ and $w_a$ are uncorrelated.
Furthermore the errors on $w_{\rm pivot}$ are the same as that on
$w_0$ with the evolution $w_a$ fixed by a prior.  To avoid confusion however,
we will refer to constraints in the context of a constant $w$ as
being on the quantity
\begin{equation}
w_{\rm DE} = w(a)= {\rm const.}
\end{equation}
The distinction between $w_{\rm pivot}$ and $w_{\rm DE}$ arises when considering
multidimensional constraints. For example in the 2D ($\Omega_{\rm DE}$,$w_{\rm DE}$) plane,
errors 
in the $\Omega_{\rm DE}$ direction increase due to marginalization over $w_a$.

\section{CMB Standards}
\label{sec:CMB}
\newcommand{\trat}{T_{\rm rat}}

CMB standards for the dark energy are particularly useful in that they
are internally or self-calibrated.  Equally importantly, the physical processes
governing these standards are sufficiently simple that there are only a few assumptions going
into the calibration.  These assumptions can themselves be tested through 
internal consistency checks.  

These standards are mainly based on the acoustic features in the power
spectrum of CMB anisotropies.   These features are frozen in place at recombination. 
The theoretical accuracy to which we can calibrate these standards is limited by
the accuracy to which recombination has been calculated.

\subsection{Recombination Aside} 

The accuracy of CMB anisotropy calculations 
are limited not by that of the radiative transfer or the general relativistic
description but ironically by that of the recombination calculation.  This status has
been true at least since the development of the modern Einstein-Boltzmann codes in
 the mid 1980's when radiative transfer improvements began to outpace recombination improvements.   The currently claimed $0.1\%$ precision of 
numerical codes (\cite{SelSugWhiZal03} 2003) is just that: a statement of precision.  
The accuracy of the standard recombination calculation has only been certified to the $\sim 1\%$ level.

The current standard for calculating the ionization history is RECFAST (\cite{SeaSasSco99} 1999)
which employs the traditional two-level atom calculation of \cite{Pee68} (1968) but
alters the hydrogen case $B$ recombination rate $\alpha_{B}$ to fit the results of 
a multilevel atom.   More specifically, RECFAST solves a coupled system of
equations for the ionization fraction $x_i$ in singly ionized hydrogen and helium ($i=$ H, He)
\begin{equation}
{d x_i \over d\ln a} = {\alpha_B C_i n_H \over H} \left[ s(x_{\rm max}-x_i) - x_i x_e \right]\,,
\end{equation}
where $n_H = (1-Y_p)n_b$ is the total hydrogen number density accounting for the
helium mass fraction $Y_p$,
$x_{e} \equiv n_e/n_H =\sum x_i$ is the total ionization fraction,  $n_e$ is the 
free electron density, $x_{\rm max}$ is the maximum $x_i$ achieved through full
ionization, 
\begin{eqnarray}
s = {\beta \over n_H} e^{-B_{1s} /k_B  T_b}\,, \quad
C_i^{-1}  =  1 + {\beta \alpha_B e^{-B_{2s}/k_B T_b} \over \Lambda_\alpha + \Lambda_{2s1s} }\,,\quad
\beta  =g_{\rm rat} \left( { k_B T_b m_e \over 2\pi \hbar^2 } \right)^{3/2} \,,
\end{eqnarray}
with $g_{\rm rat}$ the ratio of statistical weights,
$T_b$  the baryon temperature, $B_{L}$ the binding energy of the $L$th level,
$\Lambda_\alpha$ the rate of redshifting out of the Lyman-$\alpha$ line 
corrected for the energy difference
between the $2s$ and $2p$ states
\begin{equation}
\Lambda_\alpha = {1 \over \pi^2} \left( { B_{1s}-B_{2p} \over \hbar c }\right)^3 e^{-(B_{2s}-B_{2p})/k_B T_b} 
{H \over (x_{\rm max} - x_i) n_H} 
\end{equation}
and $\Lambda_{2s1s}$ as the rate for the 2 photon $2s-1s$ transition. For reference, for hydrogen $B_{1s} = 13.598$eV, $B_{2s}=B_{2p}=B_{1s}/4$,
$\Lambda_{2s1s}=8.22458 s^{-1}$, $g_{\rm rat}=1$, $x_{\rm max}=1$.  For helium
$B_{1s}= 24.583$eV, $B_{2s}=3.967$eV, $B_{2p}=3.366$eV,  $\Lambda_{2s1s}=51.3 s^{-1}$,
$g_{\rm rat}=4$, $x_{\rm max}=Y_p/[4(1-Y_p)]$. 

\begin{figure}[t]
\plotone{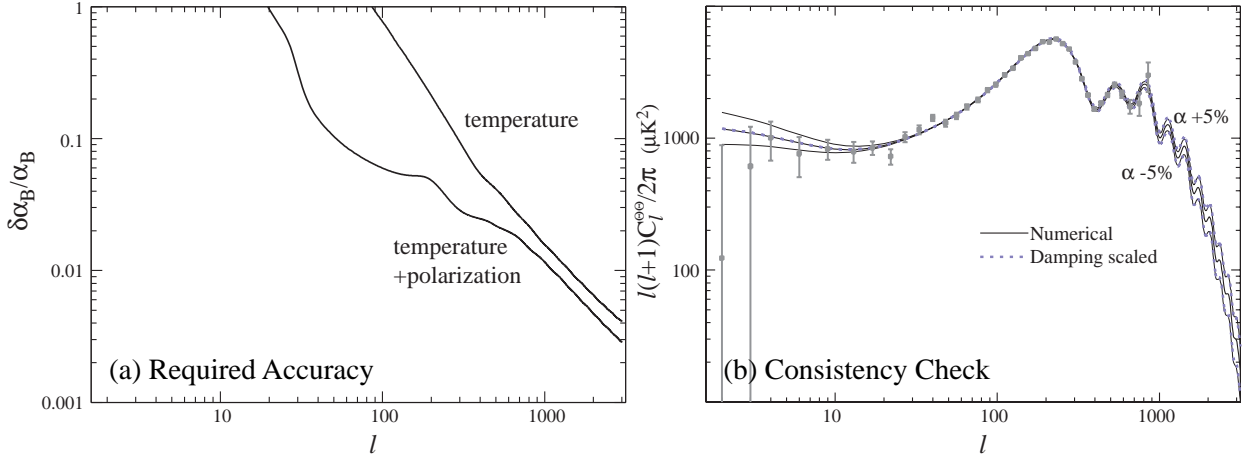}
\caption{\label{fig:recombination} Recombination and the accuracy of CMB calibrations.  (a) The current state-of-the-art in recombination involves a calibrated fudge in rescaling $\alpha_{B}$ for
hydrogen in a 2 level atom to a multilevel atom.  
Shown is precision to which $\alpha_{B}$ will need to be
calibrated so as to not introduce systematic errors that are larger than the
cosmic variance out to a given $\ell$.  (b) The damping tail contains a consistency
check for recombination, here illustrated through
 a 5\% variation in the fine structure constant $\alpha$  or a $\sim 10\%$ variation in $z_*$.
 The first 3 peaks
measured by WMAP (points) determines
the photon-baryon ratio $R_{*}$ and radiation matter ratio $r_{*}$ 
at recombination, here held fixed.  The damping tail breaks the degeneracy and measures
$z_*$ independently, here accurately analytically modeled through a change in the damping scale 
(dashed lines). }
\end{figure}

If $\alpha_B C_i n_H/ H \gg 1$, the $x_i$ reaches the Saha 
equilibrium, $s( x_{\rm max}-x_i) = x_i x_e$ or
\begin{eqnarray}
x_i &=& {1 \over 2} \left[ \sqrt{ (x_{ei}+s)^2 + 4 s x_{\rm max}} - ({x_{ei} + s}) \right] \\
   &=& x_{\rm max} \left[  1 - {x_{ei} + x_{\rm max}  \over s} \left(  1 - { x_{ei} + 2 x_{\rm max} \over s } \right) 
   + ...\right]\,, \nonumber
\end{eqnarray}
where $x_{ei}= x_e - x_i$ is the ionization fraction excluding  the species. This solution is used in place of the integration until say $x_i/x_{\rm max} -1 = 10^{-3}$.  
The recombination of hydrogenic doubly ionized helium is handled purely through the
Saha equation with a binding energy of ${1/4}$ the $B_{1s}$ of hydrogen and
$x_{\rm max}=Y_p/[4(1-Y_p)]$.  The case $B$ recombination coefficients as a function of $T_b$
are given in  \cite{SeaSasSco99} (1999) as is the strong thermal coupling between $T_b$ and
$T_{\rm CMB}$.  The multilevel-atom fudge that RECFAST introduces
is to replace the hydrogen $\alpha_B \rightarrow 1.14 \alpha_B$ independently of cosmology.

While this fudge suffices for the current observations, the recombination standard
will require improvement if CMB anisotropy constraints are to reach their full potential.
To estimate at what point the recombination calculation will need to be improved
we can compare the sensitivity of the spectra to the $\alpha_{B}$ fudge
to cosmic variance errors.  Fig.~\ref{fig:recombination} shows that at $\ell > 1000$
a calibration to the multilevel atom
that is better than the current 1\% level in $\alpha_{B}$ will be required.
Furthermore there is no guarantee that the $\alpha_B$ fudge will work to this level
of precision.

Fortunately the higher acoustic peaks provide a built in self-consistency test of
recombination.  Since the recombination rate $\alpha_B$ has been 
downgraded from a physical parameter to a fitting parameter, a more physically
interesting way to phrase the sensitivity is to change 
recombination through the fine structure constant $\alpha$.
In the recombination calculation, all of the binding energies scale as $B \propto \alpha^2$, 
$\Lambda_{2s1s} \propto \alpha^8$ and (\cite{KapSchTur99} 1999)
\begin{equation}
\alpha_B \propto \alpha^{2 (1- \partial \ln \alpha_B/\partial \ln T_b)} \,.
\end{equation}
In addition $\sigma_{T} \propto \alpha^{2}$.

The phenomenology of the CMB is primarily governed by the redshift of recombination
$z_* = a_*^{-1}-1$ though that dependence is largely hidden in standard recombination by its insensitivity
to the usual cosmological parameters.  This insensitivity and the sensitivity to $\alpha$
follows from the fact that recombination proceeds rapidly once  $B_{1s}/k T_b$ has reached
a certain threshold.   
Defining the redshift of recombination as the epoch 
 at
which the Thomson optical depth during recombination (i.e. excluding reionization) reaches unity,
$\tau_{\rm rec}(z_{*})=1$, a fit to the recombination calculation gives\footnote{
An alternate definition of the recombination redshift is that 
of the peak of the  ``visibility function'' $\dot \tau e^{-\tau}$ (e.g. \cite{Speetal03} 2003).  In principle,
the optical depth definition is more robust to sudden changes in the ionization fraction; in practice
these two definitions
 coincide to $5 \times 10^{-4}$ in the fiducial model -- a difference which can
be accounted for by replacing $1047.5 \rightarrow 1047$ in Eqn.~(\ref{eqn:recombfit}).}
\begin{equation}
\begin{array}{rcl}
a_*^{-1} &=& 1047.5 [1 + 0.00124 (\Omega_b h^2)^{-0.738}]
        [1 + b_1 (\Omega_m h^2)^{b_2} ]  \left( {\alpha \over \alpha_{0} }
       \right)^{2.08},\\
b_1 &=& 0.0783 (\Omega_b h^2)^{-0.238} [1+39.5(\Omega_b h^2)^{0.763}]^{-1}
,\\
b_2 &=& 0.560 [1+21.1(\Omega_b h^2)^{1.81}]^{-1} \,,
\end{array}
\label{eqn:recombfit}
\end{equation}
where $\alpha_0\approx 1/137$, the fine structure constant from laboratory measurements today.
The sensitivity to cosmological parameters is weak. Around the fiducial model
of $\Omega_m h^2=0.14$ and $\Omega_bh^2=0.024$, the recombination
redshift becomes
\begin{equation}
a_*^{-1}  \approx 1089 \left({\Omega_m h^2 \over  0.14} \right)^{0.0105} 
				\left( {\Omega_b h^2 \over 0.024} \right)^{-0.028} 
				\left( {\alpha \over\alpha_{0}} \right)^{2.08} \,.
\label{eqn:zstarapp}
\end{equation}
However one cannot translate the typical CMB constraints on $z_*$, $\Omega_b h^2$ 
and $\Omega_m h^2$ into constraints on $\alpha$ since the values of all three
are determined in the context of standard recombination.
We shall now see how constraints from acoustic phenomena arise
in a general context.

\subsection{Standard Rulers}

The CMB acoustic peaks are governed by 4 physical quantities:
the photon-baryon ratio, the matter-radiation ratio,  the sound horizon, 
and the diffusion scale all
evaluated at the epoch of recombination.    In the standard thermal history context,
the photon and neutrino densities are fixed by the measurement
of the CMB temperature and $T_\nu = (4/11)^{1/3} T_{\rm CMB}$. 
Defining $\trat = T_{\rm CMB}/2.725$K, the photon-baryon ratio 
becomes
\begin{equation}
R_{*} \equiv {3 \over 4}{\rho_b \over \rho_\gamma} \Big|_{a_*}=  0.729 \left( {  \Omega_{b} h^{2} \over 0.024 } \right) 
\left( {a_* \over 10^{-3} }\right) \trat^{-4} \,,
\end{equation}
and the radiation-matter ratio becomes
\begin{equation}
r_{*} \equiv {\rho_r \over \rho_m} \Big|_{a_*}= 0.297 \left( {\Omega_{m}h^{2} \over 0.14 } \right)^{-1} 
\left( {a_{*} \over 10^{-3} }\right)^{-1} \trat^{4} \,.
\end{equation}
Moreover for standard recombination $z_*$ itself is only a weak function of $\Omega_b h^2$
and $\Omega_m h^2$ [see Eqn.~(\ref{eqn:zstarapp})] and so a measurement of $R_*$ and $r_*$ translate
directly into a measurement of $\Omega_b h^2$ and $\Omega_m h^2$.  Conversely,
constraints on these two quantities 
can be generalized to  a broader context with these relations.  In Fig.~\ref{fig:recombination}b the result of 5\% variations in $\alpha$ and hence $\sim 10\%$
variations in the redshift of recombination are shown.  The variations are taken at fixed $R_*$ and
$r_*$ and hence $\Omega_b h^2$ and $\Omega_m h^2$ that vary by $\sim 10\%$.  In the
region of the first 3 peaks, these models produce nearly identical spectra.  Hence
in the general context it is $R_*$ and $r_*$ that the CMB measures not 
$\Omega_b h^2$ and $\Omega_m h^2$ directly.

 In terms of the
physics of the peaks, $R_*$ controls the baryon loading of the fluid and $r_*$
controls the depths of the gravitational potentials (see e.g. \cite{HuFukZalTeg00} 2000).  
In terms of the phenomenology, baryon loading in gravitational potential wells
modulates the relative heights of the odd and even numbered peaks whereas
the depths of the gravitational potential wells themselves also control the amplitude
envelope of all of the peaks.

With measurements of $R_*$ and $r_*$ from the morphology of the peaks, the overall
physical scale associated with the acoustic phenomena is self-calibrated under
standard recombination.  This is
the distance sound can travel by recombination
\begin{eqnarray}
s_* &=& \int_{0}^{a_*} {d a \over a^2 H(a) }c_s(a) \nonumber\\
       &=& {2 \sqrt{3} \over 3} \sqrt{ a_* \over R_* \Omega_m H_0^2}  \ln
        {\sqrt{1+R_*} +  \sqrt{R_* + r_* R_*}
        \over  1 + \sqrt{r_* R_*}}  \qquad {\rm MD/RD}
\end{eqnarray} 
where the sound speed $c_s = 1/ \sqrt{3(1+R) }$ and the second line assumes that only
matter and radiation are important in $H(a)$  before recombination.  Around the fiducial
model,
\begin{equation}
{s_* \over {\rm Mpc}}  \approx 144.4 \left( {\alpha \over \alpha_0} \right)^{-1.36} 
\left( {\Omega_m h^2 \over 0.14} \right)^{-0.252} \left( {\Omega_b h^2 \over 0.024} \right)^{-0.083}\,.
\label{eqn:sscaling}
\end{equation}
 Therefore, in the standard recombination context where $\alpha=\alpha_0$, 
 the CMB provides an absolutely
calibrated (in Mpc, not $h^{-1}$ Mpc) 
standard ruler for cosmology.  More generally the standard ruler carries
a scaling factor of $(a_*/\Omega_m h^2)^{1/2}$ if only $R_*$ and $r_*$ are determined.
This standard thermal history 
assumption is important to bear in mind when using this standard ruler for dark energy
tests.

\begin{figure}[t]
\plotone{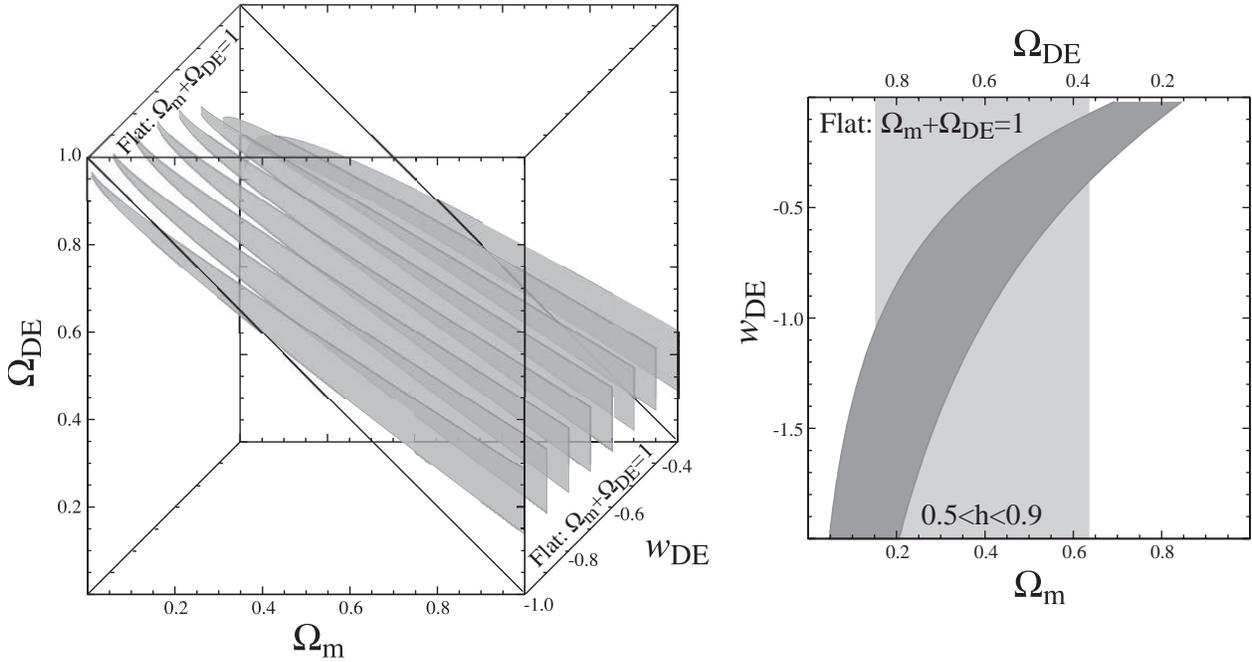}
\caption{\label{fig:cmbstuff} Generalizing the sound horizon dark energy constraint. Given CMB
 constraints on the acoustic scale $\ell_{A}$, $\Omega_{m}h^{2}$ and $\Omega_{b}h^{2}$, CMB constraints can be applied to any set of dark energy 
parameters by propagation of errors.  Shown here are WMAP constraints (a) in 
a 3D space of $\Omega_{m}$, $\Omega_{\rm DE}$, and $w_{\rm DE}$ and (b)
projected onto the usual space of flat cosmologies.}
\end{figure}

CMB constraints 
implicitly use this standard ruler in a distance measure test for
dark energy.   The sound horizon sets the fundamental physical scale of the
peaks which is then measured 
in angular or multipole space as
\begin{equation}
\ell_A = {\pi {\cal D}_* \over s_*}\,.
\end{equation}
With the absolute calibration of $s_*$, the CMB then measures the angular
diameter distance ${\cal D}_*$ to recombination in absolute units.
Note that even though the standard parameter analyses usually assume
a simple dark energy model when calculating constraints, they may
be readily translated into a general context by error propagation.
In Fig.~\ref{fig:cmbstuff} we show an example where WMAP errors
on $\ell_{A}=299\pm 2$, $\Omega_{m}h^{2}=0.14 \pm0.02$ and $\Omega_{b} h^{2}
= 0.024 \pm 0.002$ (\cite{Speetal03} 2003) are used to 
delimit a region in the three dimensional distance measure
space of curvature, constant $w_{\rm DE}$ and $\Omega_{\rm DE}$.

For quick estimation purposes or a poor-man's MCMC,  note that the dominant source of error in the
sound horizon calibration is from $\Omega_m h^2$ and that the measurement errors on
$\ell_A$ are comparatively small.   
(In fact ever since the first crude detection of the first peak in 1999 errors in the dark
energy--curvature domain have been dominated by those in $\Omega_{m} h^{2}$ 
and not by the precision with which $\ell_{A}$ was measured.  Before self-calibration became
available from the higher peaks this was phrased in terms of the prior on $h$.)
As a rule of thumb, Eqn.~(\ref{eqn:sscaling}) implies
 the errors on the distance to recombination scale as
\begin{equation}
\sigma(\ln {\cal D}_{*}) \approx {1\over 4} \sigma(\ln \Omega_{m}h^{2})\,.
\end{equation}
For example WMAP quotes errors of 14\% in $\Omega_{m}h^{2}$ and $3.6\%$ in ${\cal D}_{*}$
(\cite{Speetal03} 2003).
Improving the distance measure and the calibration of the standard ruler $s_{*}$ thus
is best achieved by accurately measuring the 3rd and higher acoustic peaks.

Beyond the third peak in the temperature, the CMB can test the
internal consistency of the standard thermal history.  The damping of the acoustic
peaks is associated with the radiative transfer or diffusion of the CMB photons during
recombination.  Under standard recombination, the diffusion distance is uniquely
determined by the baryon and matter densities $\Omega_{b} h^{2}$ and $\Omega_{m}h^{2}$
and presents another absolutely calibrated standard ruler.  However a change in 
the thermal history that delays recombination will allow the photons to diffuse
further (\cite{PeeSeaHu00} 2000)  even if the dynamics in the tight coupling regime is held fixed by keeping
$R_{*}$ and $r_{*}$ the same.  In Fig.~\ref{fig:recombination}b we see that models
with a recombination variation due to $\alpha$
start to differ at the third peak.  

Given a thermal history, the damping effect can be quantified via Boltzmann techniques
given in \cite{HuWhi97a} (1997).  The result is that the damping scale can be approximated
by
\begin{eqnarray}
\lambda_{D} = 15.96{a_{*}^{1.1} R_{*}^{-0.23} \over \sqrt{ \Omega_{m}^{0.78} \Omega_b^{0.22} H_{0}^{2} } }
(1+1.105 R_{*}^{1.87})^{1/5} [(1+r_{*})^{1/2} - r_{*}^{1/2}]^{1/2} \,.
\end{eqnarray}
In comparison to the acoustic scale, the damping scale has a stronger dependence on
the redshift of recombination and the baryons. Around the fiducial model it is approximately
\begin{equation}
{\lambda_D \over {\rm Mpc} }\approx 64.5 \left( {\alpha \over \alpha_0 } \right)^{-7/3} 
\left( {\Omega_m h^2 \over 0.14} \right)^{-0.278} \left( {\Omega_b h^2 \over 0.024} \right)^{-0.18}\,.
\end{equation} 
This scale is also projected onto 
angle via the angular diameter distance
\begin{equation}
\ell_{D} = {2\pi {\cal D}_{*} \over \lambda_{D}}
\end{equation}
and appears in the power spectrum as a sharp damping of the acoustic amplitude
by 
\begin{equation}
{\cal D}_{\ell} \approx \exp[ -(\ell/\ell_{D})^{1.25} ]\,.
\end{equation}
The ratio of the two scales $\ell_D/\ell_A$ (essentially the number of observable peaks)
is independent of the distance and dark energy and hence tests the assumptions
entering into the calibration of the standard rulers.

In Fig.~\ref{fig:recombination}b (dashed lines) we rescale the power spectrum of
the fiducial model to the $\ell_{D}$ of the varying $\alpha$ models by multiplying
by the ratio of ${\cal D}_{\ell}^{2}$.  The good agreement
explicitly demonstrates that aside from damping, the morphology of the acoustic peaks
is preserved at fixed $R_{*}$ and $r_{*}$.   

In summary the damping scale and the associated polarization is a consistency
check on the standard thermal history assumptions.  Under these assumptions both
are uniquely predicted by the low order temperature peaks.  
Problems with the assumptions
on the recombination history, e.g. the fine structure constant or a more prosaic problem
with
the  multilevel hydrogen
 atom would manifest themselves here.  Likewise problems in the assumption
of the radiation density, e.g. a change in the number or temperature of the neutrinos, would
show up here due a change in the damping scale relative to the acoustic scale and also 
due to effects from their anisotropic stress.   Finally any contamination from secondary
anisotropies and point sources would show up more strongly here.
Current small scale anisotropy and polarization
 data is in good
agreement with the predictions from the first three peaks.  These consistency tests provide
more confidence that the sound horizon is both an internally calibrated and internally
consistent standard ruler that can be used for dark energy probes, at least to the current
level of precision in the standard.

\subsection{Standard Fluctuations}

The CMB also provides a calibration standard for dark energy and/or massive
neutrino growth rate tests.  The height of the acoustic peaks is now very well
calibrated (\cite{Pagetal03} 2003) and in conjunction with the well understood radiation
transfer described in the previous sections, it can be converted into a measurement
of the initial amplitude of fluctuations.
  
For illustrative purposes, we will here assume that the neutrino masses
are negligible.   The initial spectrum of curvature fluctuations $\zeta$
\begin{equation}
\Delta_{\zeta}^{2}\equiv {k^{3} \over 2\pi^{2}}P_{\zeta}(k,a_{i})
=\delta_{\zeta}^{2} \left( {k \over k_{n}}\right)^{n-1}\,,
\label{eqn:initialpower}
\label{eqn:initialconditions}
\end{equation}
is processed through the radiation transfer to become the CMB power
spectra and matter power
spectrum. Here $k_n$ is the normalization scale 
and we will follow the WMAP convention of choosing
$k_{n}=0.05$ Mpc$^{-1}$.  Note that the actual pivot point or best constrained normalization point for
WMAP is closer to $k_{n}=0.02$ Mpc$^{-1}$ corresponding to the
first peak.  
The amplitude of fluctuations from WMAP is
\begin{equation}
\delta_\zeta \approx 5.07 e^{-(0.17-\tau)} \times 10^{-5}\,,
\end{equation}
or equivalently in terms of the WMAP normalization parameter
\begin{equation}
A = (1.84\delta_\zeta \times 10^4)^2 = 0.87 e^{-2(0.17-\tau)}  \,. 
\end{equation}
If taken instead as the amplitude at the pivot point, the uncertainties in
$\delta_\zeta$ come almost entirely from that in the optical depth $\tau$; with the
fiducial choice there is a small increase in the uncertainties due to the
constraint on the tilt. 
In any case the amplitude of
the initial fluctuations is known to better than 10\% at scales relevant to large-scale
structure.  This calibration
exceeds the accuracy to which the normalization of the power spectrum at the
current epoch is known.  Thus for dark energy tests involving the growth of structure,
it is already advantageous to compare high redshift structure to the CMB normalization
instead of the local normalization.  In addition, a measurement of the local $z=0$ normalization
also becomes a test of the dark energy.

The matter power spectrum is processed through the same radiation transfer as
the CMB and so the latter determines the former.  In the standard cosmology
this is reflected in the fact that the transfer function when expressed in Mpc$^{-1}$ depends only
on the well determined $\Omega_{b}h^{2}$ and $\Omega_{m}h^{2}$.
Thus both the shape and the normalization of the matter power spectrum in the
deceleration epoch is fixed in physical units of Mpc. 

The amplitude at $z=0$ then becomes a measure of the dark energy.  
It is usually quoted in terms of the rms of the linear
density field smoothed by a tophat of radius $r=8 h^{-1}$Mpc (\cite{HuJai03} 2003)
\begin{eqnarray}
\sigma_8^2 &\equiv& \int {dk \over k} \Delta_m^2(k,a=1) W_\sigma^2(kr)\nonumber\\
\sigma_{8} &\approx& {\delta_{\zeta} \over 5.59\times10^{-5}} 
\left( { \Omega_{b}h^{2} \over 0.024} \right)^{-1/3}
\left( { \Omega_{m}h^{2} \over 0.14} \right)^{0.563}\nonumber\\
&&\times(3.123h)^{(n-1)/2} \left( { h \over 0.72} \right)^{0.693}  
{G_0 \over 0.76}\,,
\label{eqn:sigma8}
\end{eqnarray}
where $W_\sigma(x)= 3x^{-3}(\sin x - x \cos x)$ is the Fourier transform of
a top hat window.  $G_0=G(a=1)$ the growth function evaluated at the current epoch.
Note that because the normalization is given in $h^{-1}$ Mpc, there is
a strong scaling with the Hubble constant.  In a flat universe $\Omega_{\rm DE}
= 1-\Omega_m$ and
precise measurements of $\Omega_m h^2$ then fix 
$\Omega_{\rm DE}$ given $h$.  
We shall see in the
next section that the Hubble constant is the single most useful complement to CMB parameters
for dark energy studies.
Likewise a measurement
of $\sigma_8$ is a measurement of the specific combination of dark energy
parameters above.  At high redshift, one replaces $G_0 \rightarrow G(a)$ in equation~(\ref{eqn:sigma8})
and so the measurement of structure as a function of redshift can in principle 
measure the whole growth function.

In summary, the CMB standard fluctuation
is internally calibrated, this time by the large-angle polarization-temperature cross correlation 
through $\tau$.   The inference on $\tau$ will be more thoroughly tested once the polarization auto
power spectrum becomes available.
Furthermore  an independent internal consistency check on this calibration is provided by
the gravitational lensing signature in the CMB 
(\cite{Hu01c} 2002; \cite{KapKnoSon03} 2003).

\section{Standard Deviants}
\label{sec:deviations}

The CMB has provided  two self-calibrated
standards for dark energy studies, a standard ruler:
the sound horizon at recombination and a standard fluctuation: the initial amplitude of fluctuations
at the $k=0.05$ Mpc$^{-1}$ scale.  Their respective calibration errors are currently $<4\%$ and
$<10\%$ respectively.   With the sound horizon calibration, the peak locations constrain
the distance to the recombination surface to the same fractional accuracy.

With the addition of better small scale anisotropy
and large scale polarization measurements one can expect that the statistical
 errors
will further improve by up to a factor of 10.   Moreover internal consistency tests should 
ensure accuracy and test for systematic errors.
It is therefore interesting to reconsider the dark energy observables
keeping the CMB standards, i.e. the high redshift universe and hence ${\cal D}_*$,
$G_*$, $\Omega_b h^2$ 
and $\Omega_m h^2$ fixed.    We shall see that this perspective leads to several
counterintuitive results which highlight the importance of accurate local calibration over
redshift range or depth.

\begin{figure}[t]
\plotone{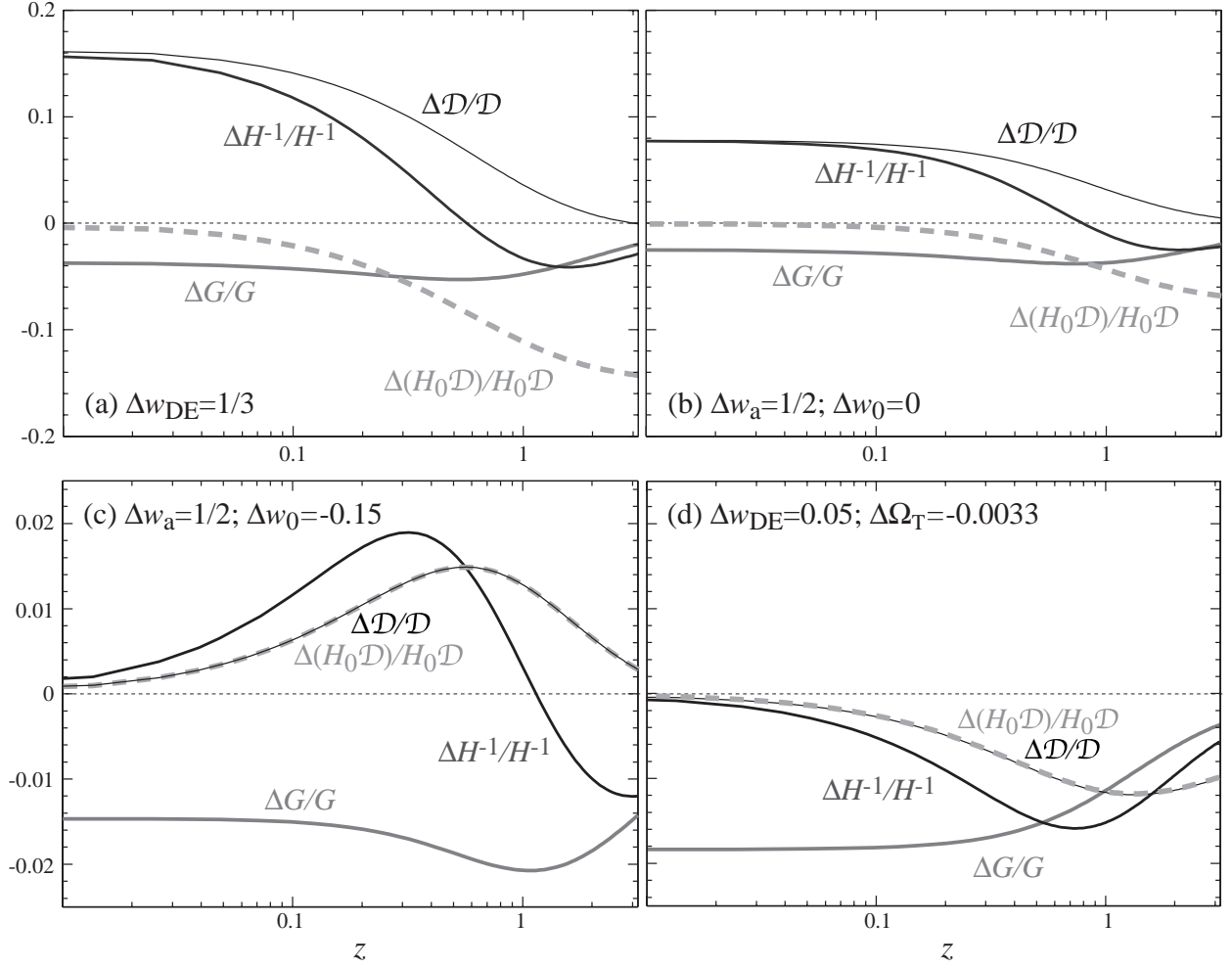}
\caption{\label{fig:deviation} Deviations in the dark energy observables holding CMB observables 
${\cal D}_{*}$ and $G_{*}$ fixed by varying $\Omega_{\rm DE}$ to compensate a
variation in (a) a constant $w_{\rm DE}$;  (b) $w_{a}=-dw/da$ at fixed $w(a=1)=w_{0}$; compensating
variations which leave $H_0$ fixed
(c)  $\Delta w_{a}/\Delta w_{0}\approx -10/3$ 
(d) $\Delta w_{\rm DE}/\Delta \Omega_{T} \approx -15$.  
With fixed high-$z$ observables, the main deviations due to the
dark energy equation of state appear as variations in the Hubble constant which can 
be measured at low redshift by absolute standards through ${\cal D}$, $H$ or at high
redshift through relative standards $H_{0}{\cal D}$.  The local value of the growth function
$G_0=G(a=1)$ is useful 
in breaking the degeneracy left by variations at fixed $H_{0}$.  }
\end{figure}

In Fig.~\ref{fig:deviation}a we show deviations in the distance and growth dark energy observables
from a fiducial model of $\Omega_{\rm DE}=1-\Omega_m=0.73$ and $w_{\rm DE}=-1$
(with $\Omega_{m}h^{2}=0.14$ and $\Omega_{b}h^{2}=0.024$).   
These are the angular diameter distance ${\cal D}$, the Hubble parameter $H$, the growth function
$G$ and the combination $H_0 {\cal D}$, the distance as measured by a comparison of local
and high redshift standards.  Note that with $\Omega_m h^2$ fixed, ${\cal D}$ and $H$ are
known functions of redshift in the matter dominated or deceleration regime.

Let us begin with a variation of $\Delta w_{\rm DE} = 1/3$, i.e. a model with $w_{\rm DE}=-2/3$. 
Given a fixed ${\cal D}_*$, $\Omega_{\rm DE}$ is then not a free parameter
but fixed given $w_{\rm DE}$.   This is similar to but {\it not} equivalent to the usual
fixing of $\Omega_m$ or $\Omega_{\rm DE}$ by a prior found in SNe forecasts.  The CMB prior on ${\cal D}_{*}$
is simple to apply
and should be used in any projection involving the CMB (e.g. \cite{HuEisTegWhi99} 1999; \cite{SpeSta02} 2002; \cite{FriHutLinTur03} 2003).

Fixing ${\cal D}_*$ changes the perspective on where in redshift the largest
deviations from the fiducial $\Lambda$ model would appear.   Not surprisingly given that
the CMB is a high redshift probe, in terms
of absolute distances and growth, it is $z \rightarrow 0$ where most of the effects are largest.
In fact the single most useful measurement that would complement the CMB distance
measure is a Hubble constant measurement that is accurate to the percent level.  Note
that at $z\rightarrow 0$ deviations in the Hubble parameter $\Delta H/H$ and the angular diameter distance $-\Delta {\cal D}/ {\cal D}$ are given by that in the Hubble
constant $\Delta H_0/H_0$.   Likewise, since the dark energy only affects the growth
of structure during the acceleration epoch, the deviations in it
also appear only at $z \simlt 1$.  Here nearly compensating variations in $w_{\rm DE}$
and $\Omega_{\rm DE}$ imposed by ${\cal D}_*$ keep the fractional variations in $G$
small compared with those in $H_0$ but recall that redshift survey based measurements
of fluctuations carry a strong scaling with $H_0$ on top of the growth rate
[see Eqn.~(\ref{eqn:sigma8})].

Relative standards that measure $H_0 {\cal D}$ are an exception to the low redshift
rule but not an exception to the $H_0$ rule.   Here the maximum deviation appear
at high-$z$ but since ${\cal D}$ is fixed at high-$z$, $\Delta H_0 {\cal D}/ H_0 {\cal D} 
\rightarrow \Delta H_0/H_0$.  In other words, since the distance to high redshift is
known, measurement of a standard candle becomes a calibration of the absolute
brightness of the standard.   Phrased another way, relative standards measure
$\Omega_m$ in the deceleration regime and when combined with $\Omega_m h^2$ from the
CMB determine $H_0$. The CMB provides a counter-intuitive way of measuring
the Hubble constant by inverting the distance ladder!   Of course in practice,
the assumption that the candle is standard or standardize-able between redshift zero
and the deceleration epoch is suspect.  Note also that Fig.~\ref{fig:deviation}a only
shows the best epoch to measure the dark energy observables given fixed fractional
measurement errors in the observables.  It does not factor in the observational cost
required to achieve a fixed fractional distance error at high redshift.

These rules of thumb remain valid for deviations involving an evolution in the equation of
state from its present value.  In Fig.~\ref{fig:deviation}b we show the deviations for $\Delta w_a=1/2$
with $w_0=-1$.    Again the same statements apply: the maximal deviations for absolute
standards appear at redshift zero and for relative standards at high redshift.  For the
distance measures, the asymptotic deviations correspond to variations in the Hubble constant.  

The deviations due to $w_{0}$ at fixed $w_a=0$ (or equivalently $w_{\rm DE}$) and those
for $w_a$ at fixed $w_0$ have similar forms since to leading order they are both tied
to uncertainties in the Hubble constant.  This
similarity implies a degeneracy between $w_0$ and $w_a$ along a line of 
$\Delta w_0 / \Delta w_a \approx -3/10$ which holds $H_0$ fixed.  
As noted in \S \ref{sec:parameterization}, a degeneracy between $w_0$ and $w_a$
simply implies that the equation of state  is better constrained at a redshift that differs
from $z=0$.  In this CMB context with fixed high-$z$ quantities, this implies $a_{\rm pivot} 
= 7/10$ from Eqn.~(\ref{eqn:pivot})
or $z_{\rm pivot}=0.43$ as the epoch at which the dark energy equation of
state is best constrained.
Recall that a measurement of $w \ne 1$ at any redshift
would rule out a cosmological constant.    Conversely a confirmation of $w = -1$ at such
a redshift would strongly favor  a cosmological constant since an alternate solution would
require a $w_a$ variation that sent the dark energy to a phantom regime $w<-1$ in the past 
expansion history, or a stronger variation in $w$ that violated the $w_a$ linearization.

In Fig.~\ref{fig:deviation}c, we plot the
deviations along this degeneracy line (deviations are here actually between $w_0=-0.85$, $w_a=0$
and $w_0 = -1$, $w_a=1/2$ to avoid the phantom dark energy regime).   Note the change in scale of the axes.  Because this line preserves the Hubble constant, the deviations in ${\cal D}$ and $H_0 {\cal D}$
now coincide and disappear at both low and high redshift.   The Hubble parameter deviations persist until higher redshift and the growth function deviations remain
fairly level for $z\simlt 3$.  Note that still a measurement of the growth function locally provides
a means
of breaking the degeneracy.  In any case, any means of breaking the degeneracy left at constant
$H_0$ will require percent level accuracy in the measurements and calibration of the standards.

Finally these patterns  also appear with variations involving the curvature or $\Omega_{T}$.
As is well known, variations in the curvature off 
of a flat cosmology are allowed by the CMB but 
come at the cost of a change in the Hubble constant.  Again an accurate Hubble
constant is the key.  With $w_{\rm DE}$, a more subtle degeneracy with $\Omega_{T}$ at
fixed $H_{0}$ opens up (see Fig.~\ref{fig:deviation}d).  
In this case measurement of the growth function at the present epoch becomes even
more important since the distance degenerate increment in $w_{\rm DE}$ and 
decrement in $\Omega_{T}$ both slow the growth of structure.  Note though that at the
$1\%$ level in growth rate, massive neutrinos will certainly have to be included in the interpretation
and error budget.

In summary, to test the cosmological constant hypothesis and measure the equation of state
of the dark energy at $z\sim 0.4-0.5$, the best complement to current 
and future CMB
measurements is a measurement of the Hubble constant that is accurate at the few
percent level.  Ironically, one
way of achieving this is to measure the relative luminosity distance to a redshift in 
 the deceleration regime.  If the measurement is inconsistent with a cosmological constant, then
to further measure the evolution of $w$ through $w_a$ or rule out an alternate
explanation involving spatial curvature
will require percent level measurement
and calibration of standard candles, rulers, number densities at intermediate redshifts
or fluctuations at any redshift. 

\section{Standard Forecasts?}
\label{sec:forecasts}

The general considerations of the previous sections
 can be turned into specific forecasts for dark energy parameters
given assumptions about the observations in question: both on the CMB side and
on the dark energy probe side.    If the observations are expected to constrain the models to live
in a small region of parameter space and systematic errors are negligible, then linear propagation 
of statistical errors, also known as
Fisher matrix forecasts, provide a useful guide to their capabilities.  

Unfortunately one can make a dark energy parameter forecast 
give practically any desired answer by adjusting prior assumptions both on the cosmology
and the systematic error floor.
(First rule of parameter estimation: state your priors.  Second rule of parameter estimation: 
state your priors.)   As an exercise, here we will try to compare on an equal footing (when titles end with a question mark, the answer is no)
several different dark energy probes that can come out of a deep and wide optical survey.
Specifically we assume a multi-color survey that allows binning of galaxies
to $\Delta z = 0.1$ out to  $z=1$ across
4000 deg$^2$.
We will assume CMB priors that come from a
forecast of the Planck satellite assuming only statistical errors in
a parameter space that includes $\Omega_{m}h^{2}$, $\Omega_{b}h^{2}$, $n$,
$\delta_{\zeta}$, tensors, and $\tau$ as well as the dark energy parameters of interest (\cite{Hu01c} 2002).  
As we 
have seen in \S \ref{sec:CMB}, the critical CMB assumptions are on the physical matter density
$\sigma(\ln \Omega_m h^2)=0.01$ which controls the calibration of the sound horizon and $\sigma(\tau)=0.005$ which controls the calibration of the growth function.  We also assume
a spatially flat universe.

\begin{figure}[t]
\plotone{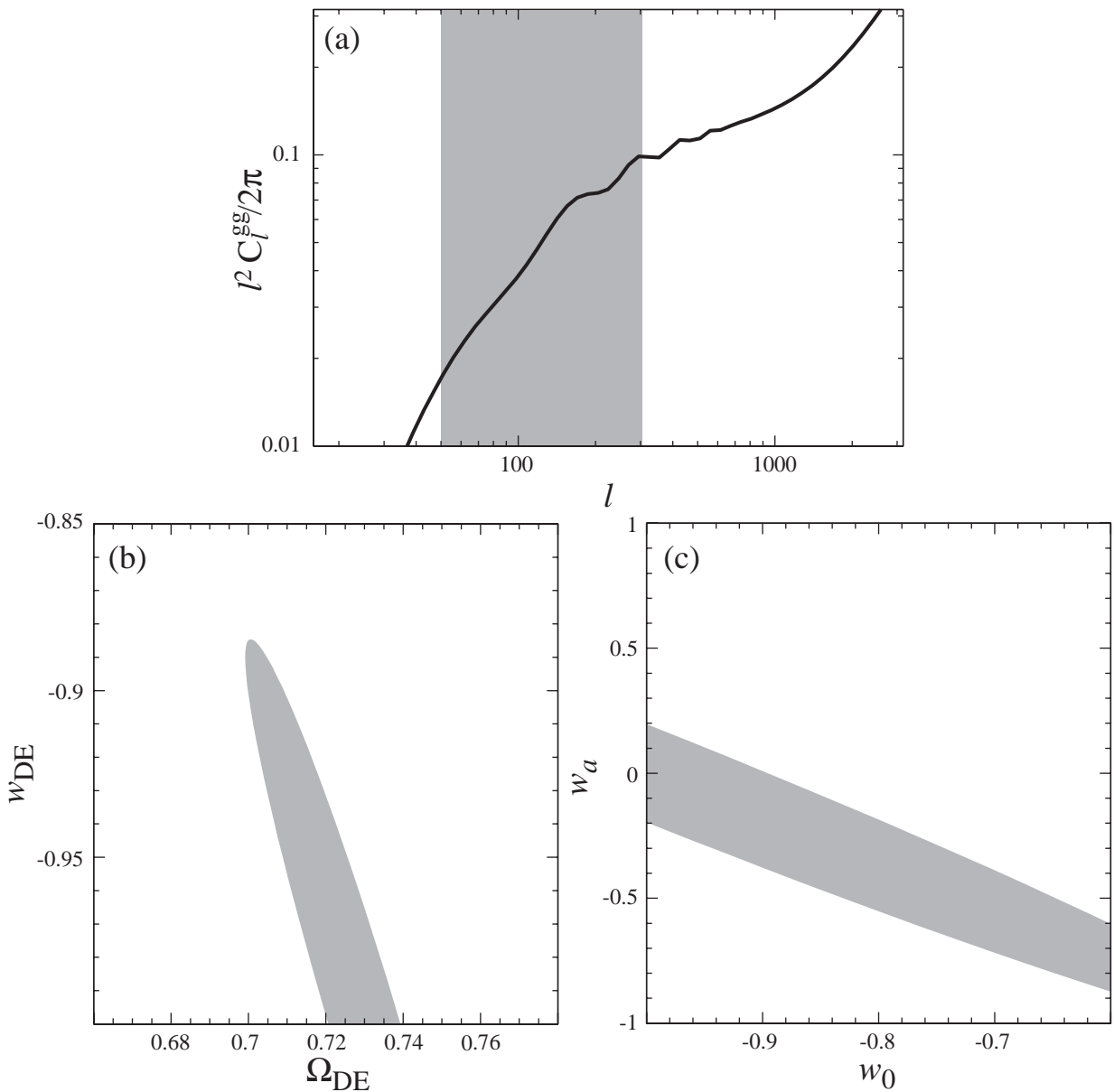} 
\caption{\label{fig:wiggles} Acoustic or baryon features in the galaxy angular power spectrum.  (a)
angular power spectrum at $z=1$ with $\Delta z=0.1$
 for galaxies in halos above $M_{\rm th} = 10^{12.5} h^{-1} M_\odot$; (b) constraints on $w_{\rm DE}$ and $\Omega_{\rm DE}$ from $50 \le \ell \le 300$ and 10 angular spectra out to $z=1$ with Planck CMB priors; (c) 
 constraints in the $w_{0}-w_{a}$ plane marginalized over $\Omega_{\rm DE}$.}
\end{figure}

Given that 
 the sound horizon at recombination is already calibrated at the few percent level, its appearance in the matter power spectrum
as baryon features or ``wiggles" provides a theoretically robust, but observationally
challenging, probe of the dark energy (\cite{EisHuTeg99a} 1999).  
Much recent work has gone into their utilization in
a high-$z$ redshift survey (e.g. \cite{BlaGla03} 2003; \cite{HuHai03} 2003; \cite{SeoEis03} 2003).

It is interesting to investigate to what extent a photometric survey can utilize
the baryon features.  A photometric survey essentially loses all
ability to measure the Hubble parameter $H(a)$ directly
through the standard ruler in the radial direction unless $\Delta z \ll 0.01$.   As a means
of measuring the angular diameter distance ${\cal D}$, a photometric survey fairs better
(\cite{CooHuHutJof01} 2001).
At high redshift, the change in the distance per unit redshift is small and
even a fairly thick shell of $\Delta z=0.1$ produces very little smearing of the features due to
projection.  In Fig. \ref{fig:wiggles}, we show the angular power spectrum of galaxies in haloes
$M\ge M_{\rm th}= 10^{12.5} h^{-1} M_\odot$ predicted under the halo model described in
\cite{HuJai03} (2003).  Of course, a measurement of ${\cal D}$ from the angular power spectrum still
does not rival that from the 3D power spectrum given the loss of radially directed modes.
The relative degradation as a function of redshift resolution can be estimated by a simple
mode counting argument (\cite{HuHai03} 2003).

Still the angular power spectrum does yield
 some constraint on the
dark energy.   In Fig.~\ref{fig:wiggles}, we show the results of a Fisher forecast with 10 
redshift bins $z\le 1$  taking only the quasi-linear regime $50 \le \ell \le 300$ for the where the halo
model for the galaxy clustering is relatively robust.  To be conservative we marginalize 
over 5 halo model parameters per redshift bin as described in  \cite{HuJai03} (2003).
In the context of a constant $w_{\rm DE}$, angular features allow a joint
determination  with $\Omega_{\rm DE}$.
In the context of the $w_0-w_a$ plane (marginalized over
$\Omega_{\rm DE}$) there remains a degeneracy  that lies close to a line of constant $H_0$.
Although these constraints are relatively weak, they are theoretically robust and come more
or less for free given an optical survey with a well quantified selection.

An optical galaxy survey can also exploit the CMB standard fluctuation calibration if
one can relate the observables to the underlying mass spectrum. One way of doing
so is to also measure the gravitational lensing shear distortion of the
background galaxy images by mass in the foreground.
Consider first consider the mass associated with the dark matter halos around
foreground galaxies called galaxy-galaxy lensing.
By correlating the foreground galaxy positions with the background
galaxy shears, one measures the galaxy-shear angular power spectrum.  
Combined with the galaxy-galaxy  angular power spectrum, one can extract the bias
in the linear regime or more generally the bias divided by the galaxy-mass correlation
coefficient.  Given a halo model for the association of galaxies with dark matter halos,
even the latter can be converted into a measurement of the mass power spectrum 
(\cite{GuzSel02} 2002).

\begin{figure}[t]
\plotone{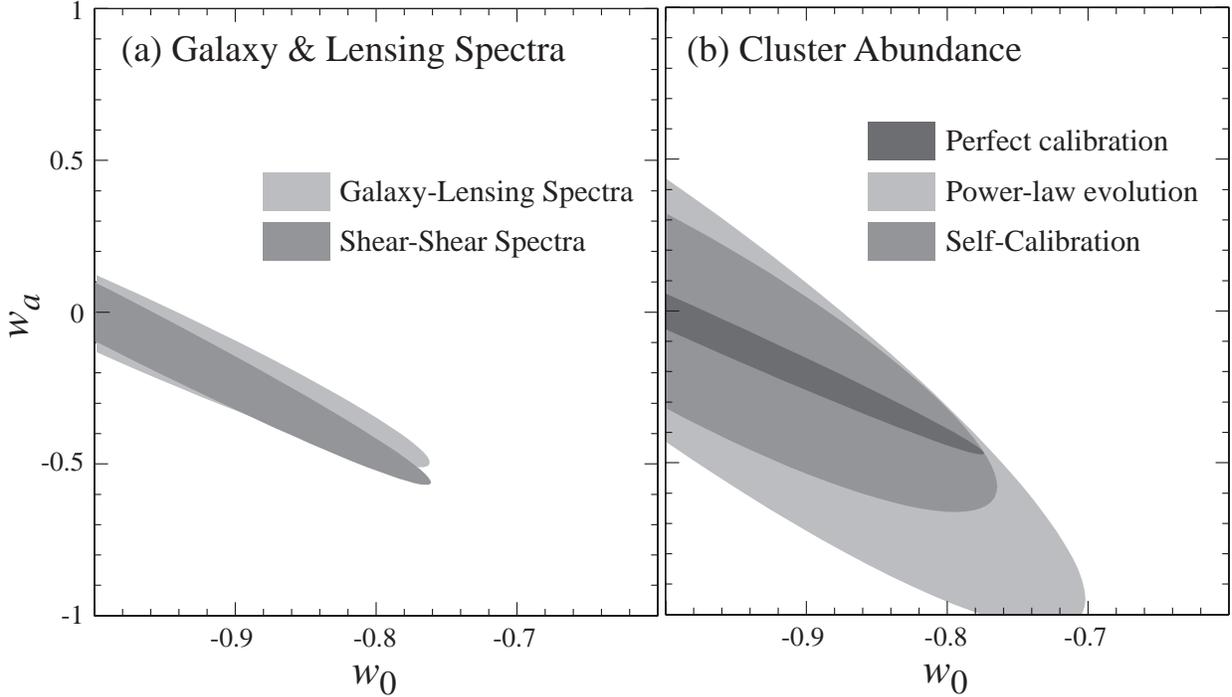}
\caption{\label{fig:shearabundance} Dark energy constraints from (a) galaxy and lensing spectra 
(b) cluster abundance.  10 galaxy bins of $\Delta z=0.1$ out to $z=1$ and 4
shear bins of $\Delta z=0.25$ out to $z=1$ plus a $z \ge 1$ bin with shear noise
corresponding to $\bar n=10$ gal arcmin$^{-2}$ and $\gamma_{\rm rms}=0.16$.  
Galaxy-galaxy
and galaxy-shear power spectra constraints from $50 \le \ell \le 1000$ are compared
with shear-shear spectra from $50 \le \ell \le 3000$.   The cluster abundance is
divided into the same bins as the galaxies for $M>  10^{14.2} h^{-1} M_\odot$ with
a comparison between perfect mass calibration, marginalization over a power law
mass-observable relation, and self-calibration through employing the sample variance
as a measure of the mass dependent clustering of clusters. }
\end{figure}

As a proof of principle let us again consider the halo model in \cite{HuJai03} (2003) but
now allow constraints from $50 \le \ell \le 1000$ from the 10 foreground galaxy redshift
bins out to $z=1$ again selecting $M_{\rm th} = 10^{12.5} h^{-1} M_\odot$.  For the 
background galaxies, we assume 4 bins of $\Delta z=0.25$ out to $z=1$ and an additional
bin for all higher redshift background galaxies.  We choose a distribution of background
galaxies with a median redshift of $z_{\rm med}=0.7$, an angular density of $\bar n = 10$ gal
arcmin$^{-2}$, and a shear error per component per galaxy of $\gamma_{\rm rms}=0.16$.
With these assumptions the combination of galaxy-shear and galaxy-galaxy power spectra 
can constrain $w_0$ and $w_a$ separately as in Fig.~\ref{fig:shearabundance}.

Galaxy-lens constraints on the dark energy require
 the assumption of a model for the association
of galaxies to the dark matter whose validity must be verified.
Shear-shear correlation on the other hand depend only on the large-scale matter
distribution and is theoretically more robust but observationally more difficult to
measure. Constraints from  the cross correlation
of the shear-shear measurements from the same binning scheme and noise assumptions as above
are shown in Fig.~\ref{fig:shearabundance} 
(\cite{Hu99b} 1999).  Here we have taken $50 \le \ell \le 3000$
since the shear field remains Gaussian and theoretically predictable
 to smaller scales than the galaxy field.  
Shear-shear and galaxy-lens data of this quality have similar constraining power
with regards to the dark energy and may be used to cross-check each other.

Finally let us consider the cluster abundance.   The cluster abundance at a fixed mass
is exponentially sensitive to the amplitude of linear fluctuations at a given redshift 
(see e.g. \cite{HaiMohHol01} 2001).   
Given the CMB normalization of fluctuations at high redshift even the local cluster abundance
becomes a constraint on the dark energy that can measure a constant equation of state
(see Fig.~\ref{fig:deviation}a).   At high redshift, the cluster abundance can measure the
evolution of the equation of state (\cite{WelBatKne01} 2001).  The central concern
is whether the cluster selection can be calibrated in mass at the required level
of accuracy of better than a few percent.  The main hope is that the wealth
of observables beyond the optical, extending from the radio to X-ray frequencies, will
allow an accurate mass calibration.
Given that the full suite will only be available at fairly low redshift,
 it may be wiser to focus on the  mass calibration of the local sample when measuring
$w_{\rm DE}$ (\cite{KunCorParCop03} 2003).  Recall also that the total growth to $z=0$ is also a good
way to separate curvature and dark energy effects.

Nonetheless, let us suppose that  
we can count
all of the clusters above $M_{\rm th} = 10^{14.2} h^{-1} M_\odot$ in the 10 redshift bins.  With
a perfect calibration of the mass threshold, the cluster abundance can provide interesting
constraints on both $w_0$ and $w_a$ with the main degeneracy line again following
a constant $H_0$ (see Fig.~\ref{fig:shearabundance}).   
If on the other hand, the mass-observable relation that controls the mass
threshold $M_{\rm th}$ is allowed to undergo a power law evolution in redshift which must
be determined by the abundance measurements themselves, the constraints
degrade substantially, especially in the $w_a$ direction.   Fortunately, clusters have
more observables then simply their abundance
above threshold in a single observable.
With multiple observables it is possible to self-calibrate the mass threshold in principle.  Here
we show a minimal example of self calibration which involves the sample variance of the cluster
counts themselves and hence is fully internal to a cluster abundance survey.   Some of the
lost information is regained since the sample variance or clustering of clusters as a function of
their mass is known from simulations (see \cite{LimHu04} 2004; \cite{MajMoh03} 2003 for details).

\section{Discussion}

The CMB has already provided a set of accurately calibrated standards for dark energy
studies.  The sound horizon at and distance to recombination is measured to better than 4\%
and the amplitude of initial fluctuations at large-scale structure scales of $k =0.05 $Mpc$^{-1}$
to better than 10\% by the WMAP data alone.  
Moreover, the expansion history, e.g. distances, volumes and the Hubble
parameter,  during the whole deceleration epoch
has been determined to the level controlled by errors in  $(\Omega_m h^2)^{1/2}$  -- currently 
less than $5\%$ once all of the CMB data is considered.  With the high-$z$ expansion history fixed,
a measurement of even local $z=0$ observables can determine the dark energy
equation of state.  The caveat to using this long lever arm to measure the dark energy
is that the local and high redshift
standards must be accurately calibrated.

CMB inferences are based on an interpretation of the acoustic peaks that has passed
internal consistency checks in the damping tail and polarization.   
The critical assumption underlying the interpretation is
the thermal history of the universe and we have focused on recombination and reionization
as a challenge for the  sub percent level calibration of CMB standards.

Given expected improvements in the measurements of CMB standards, 
deviations due to the dark energy in distance and growth measures appear mainly
at low redshift.  The former mainly represent deviations in the Hubble constant. It is
ironic that the primary quantity that dark energy probes measure in light of the CMB is
the Hubble constant.  This includes high-$z$ SNe.  Nonetheless,
a Hubble constant determination would measure the dark energy equation of
state at $z \sim 0.5$ in a flat universe at comparable fractional precision.  
If this quantity is measured to be inconsistent with a cosmological
constant, then distance measures at intermediate redshifts or growth measures at
any redshift can be used to test its evolution and/or contamination in its determination
from a small spatial curvature.  However, in this case the expected deviations are smaller
and will require even more accurate measurement and calibration of standards.

CMB standards are naturally exploited by deep optical surveys.  The sound horizon appears
as baryonic features in angular and spatial power spectra and can be used to measure distances.
The density fluctuation calibration enters into the clustering of galaxies, galaxy-galaxy lensing
and cosmic shear as well as the abundance of rich clusters.  Exploiting these standards
with observations that match the accuracy of CMB determinations will be the challenge for
future dark energy probes.

\smallskip{\it Acknowledgments:}  I thank D. Holz, E. Sheldon and 
M. White for useful discussions.  
This work was supported by the DOE and the Packard Foundation and carried out
at the KICP under NSF PHY-0114422.

\end{document}